# Characterization of the Burst Stabilization Protocol for the RR/RR CICQ Switch

Neil J. Gunther
*Performance Dynamics Company*
*4061 East Castro Valley Blvd., Suite 110*
*Castro Valley, CA 94552*
njgunther@perfdynamics.com

Kenneth J. Christensen and Kenji Yoshigoe
*Computer Science and Engineering*
*University of South Florida*
*Tampa, FL 33620*
{christen, kyoshigo}@csee.usf.edu

**Abstract**

*Input buffered switches with Virtual Output Queueing (VOQ) can be unstable when presented with unbalanced loads. Existing scheduling algorithms, including iSLIP for Input Queued (IQ) switches and Round Robin (RR) for Combined Input and Crossbar Queued (CICQ) switches, exhibit instability for some schedulable loads. We investigate the use of a queue length threshold and bursting mechanism to achieve stability without requiring internal speed-up. An analytical model is developed to prove that the burst stabilization protocol achieves stability and to predict the minimum burst value needed as a function of offered load. The analytical model is shown to have very good agreement with simulation results. These results show the advantage of the RR/RR CICQ switch as a contender for the next generation of high-speed switches.*

## 1. Introduction

High-speed switches are the core of the Internet and make possible end-to-end delivery of packets. These switches have been designed typically to have their packet buffers at the output ports. Output Queued (OQ) switches require buffer memories that are $N$ times as fast as link speed (for $N$ input ports). Link speeds are increasing much more rapidly than memory speeds [11]. In Input Queued (IQ) switches buffer memories need only match the link speed. Thus, IQ switches have been the subject of much research for high-speed switching. To overcome head-of-line blocking in IQ switches, Virtual Output Queueing (VOQ) [1, 16, 23] is employed. VOQ switches require switch matrix scheduling algorithms to find one-to-one matches between input and output ports. VOQ switches can be IQ, Combined Input and Output Queued (CIOQ), or Combined Input and Crossbar Queued (CICQ). Fig. 1 shows a generic VOQ switch. CICQ switches have limited buffering at the Cross Points (CP) and have become feasible with the continued

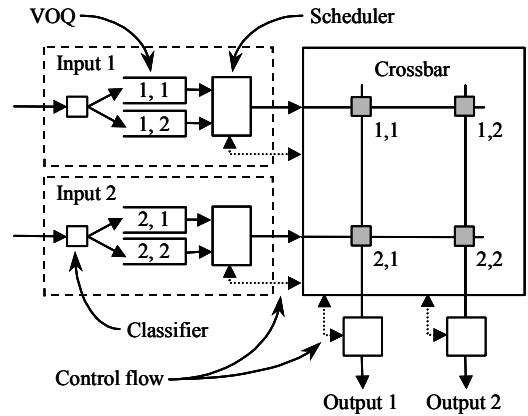

**Figure 1. VOQ input buffered switch**

increase in density of VLSI technologies [12, 19, 20, 21, 26, 27, 28].

The trade-offs in VOQ switch matrix scheduling are stability, fairness, and implementation complexity. Stability refers to bounded queue length for schedulable loads (an unstable queue has an unbounded queue length). IQ cell switches use iterative request-grant-accept scheduling cycles to achieve a maximal one-to-one matching. Existing scheduling algorithms for IQ cell switches based on an unweighted maximal matching (such as PIM [1] and iSLIP [16]) are not stable unless internal speed-up is used. A 2x speed-up has been proven to be sufficient for stability for all schedulable flows [5]. If a weighted maximal match is implemented based on Oldest Cell First (OCF) or Longest Queue First (LQF), stability can be achieved for all schedulable flows without speed-up [18]. LQF can be unfair since it can cause starvation. Weighted matching requires state information to be exchanged between input and output ports. CICQ switches can use independent round robin (RR) selection of VOQs and CP buffers (this is an RR/RR CICQ switch). However, instability occurs unless OCF or LQF is used to select VOQs in an input port [12]. Both OCF and LQF require comparisons between all $N$ ports during each scheduling cycle. This requires either $N$ sequential

---

This material is based upon work supported by the National Science Foundation under Grant No. 9875177 (Christensen and Yoshigoe).

comparisons or $Log_2(N)$ comparisons with a tree circuit containing $N-1$ comparators. Better methods for achieving stability are needed. We address this need.

In this paper, Section 2 briefly reviews VOQ switch architectures and their scheduling algorithms. Section 3 describes an unstable scheduling region in iSLIP and RR/RR CICQ switches and proposes a burst stabilization protocol to overcome this instability. Section 4 evaluates burst stabilization for the RR/RR CICQ switch and develops a model to predict the minimum burst values required to achieve stability. Section 5 is a summary.

## 2. Overview of VOQ Switches

For many years, IQ switches were considered to be an academic curiosity due to their poor performance caused by head-of-line blocking. The breakthrough in IQ switch architectures occurred when Virtual Output Queueing (VOQ) was invented by Tamir and Frazier [23] in 1988 and then developed by Anderson et al. [1] and McKeown [16] in the early 1990's for cell-based packet switching. In a VOQ IQ switch, each input buffer is partitioned into $N$ queues with one queue for each output port (hence the name "virtual" output queueing).

IQ cell switches with bufferless crossbars use iterative request-grant-accept scheduling cycles to achieve a maximal matching of inputs to outputs. In Parallel Iterated Matching (PIM) [1] the steps are:
1. Each unmatched input sends a request to every output for which it has a queued cell.
2. If an unmatched output receives any requests, it grants to one input by randomly selecting a request among those requesting to this output.
3. If an input receives a grant, it accepts one by randomly selecting an output among those granted to this output.

In iSLIP [16], accept and grant counters are maintained in each input and output port, respectively, and result in a randomized matching at high utilization due to a "slip" between counters. Iterated matching algorithms require control flow between all input and output ports. This coupling between ports adds complexity to a switch implementation.

IQ switches with limited buffering in the crossbars – CICQ switches – became feasible to implement in the late 1990's [21, 27, 28]. Scheduling in a CICQ switch can be RR for both the VOQs within the input ports and for the CP buffers within the crossbar to achieve an RR/RR CICQ switch as shown in Fig. 2. Each CP buffer needs capacity to hold only two cells. For CICQ switches, control flow can be reduced to only requiring knowledge at the input ports of crossbar buffer occupancy. CICQ switches thus uncouple input and output port scheduling and are less complex than IQ switches with bufferless crossbars. In [27] the design of a 10-Gbps RR/RR CICQ switch using off-the-shelf FPGA technology is presented.

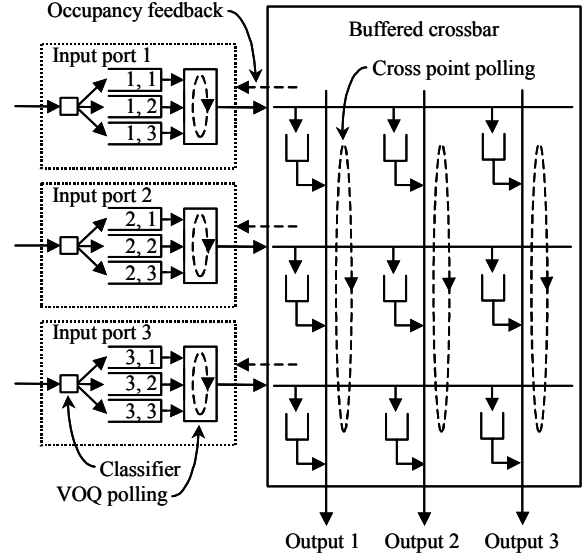

**Figure 2. RR/RR CICQ switch**

## 3. Unstable Regions in VOQ Switches

In this section, we consider an unstable region in the iSLIP and RR/RR CICQ switches. We assume that arrivals are Bernoulli with rate $\lambda_{ij}$ for $i, j = 1,\ldots,N$ where $i$ is the input port number, $j$ the output port number, and $0 \le \lambda_{ij} \le 1.0$. A Bernoulli model is a common traffic model for studying switch performance (e.g., as used in [1, 12, 16, 18]). The Bernoulli model is described in Section 4.1.

**Definition 1** *Let $\lambda_1$ be the offered load at port 1. The fraction f of offered traffic going to:*

1. *$VOQ_{11}$ is $f\lambda_1 = \lambda_{11}$*
2. *$VOQ_{12}$ is $(1-f)\lambda_1 = \lambda_{12}$*

*where $1/2 < f < 1$.*

**Corollary 1** $\lambda_1 = \lambda_{11} + \lambda_{12}$

**Remark 1** *Note that:* $\lambda_{12} \ne 1 - \lambda_{11}$ *unless* $\lambda_1 = 1$. *The mean interarrival time of cells at port 2 is* $\tau_2 = \lambda_2^{-1}$. *But* $\lambda_{21} \equiv \lambda_2$ *by virtue of* $\lambda_{22} = 0$.

In [12] a region of instability for iSLIP IQ and RR/RR CICQ switches is demonstrated for a schedulable, asymmetric traffic load to two ports. For any two ports arbitrarily identified as ports 1 and 2, let

$\lambda_1 = \lambda_{11} + \lambda_{12}$, $\lambda_{21} = \lambda_{12}$, and $\lambda_{22} = 0$. Within a region of $\lambda_{11} > 0.5$ and high offered traffic load, instability occurs. This instability condition is not limited to a two-port switch, but can occur between any two ports of a large switch. The offered load which causes instability for an RR/RR CICQ switch ranges from a low of approximately 0.9 in the range of $0.6 < \lambda_{11} < 0.7$ to a high of 1.0 at $\lambda_{11} = 0.5$ and $\lambda_{11} = 1.0$. This instability exists for a switch of size $N$ input and output ports where any two of the $N$ ports have the traffic load specified above. The instability range for an iSLIP IQ switch is larger in area. The simulation models developed and validated in [26] are used to reproduce the instability results in [12]. Infinite size VOQ buffers are assumed for both an iSLIP IQ and CICQ switch. For the iSLIP IQ switch, four iterations are used per scheduling cycle.

As an experimental means to detect instability, simulation experiments were run for 100 million cell times and terminated as unstable if any queue length exceeds 5000 cells. A similar experimental means of detecting instability is used in [8]. Fig. 3 shows the simulation results for the iSLIP and RR/RR CICQ instability regions and this exactly matches the results in [12]. For this same simulation experiment, OCF/RR and LQF/RR for a CICQ switch do not exhibit instability.

The instability in the RR/RR CICQ switch is caused when $VOQ_{12}$ is empty (drained) and $VOQ_{11}$ is blocked from transferring a cell to its cross point buffer ($CP_{11}$) due to $CP_{11}$ being already full and $CP_{21}$ transferring to output 1. A solution is to service $VOQ_{12}$ less aggressively so that $VOQ_{12}$ will have queued cells that can be transferred (to $CP_{12}$) when $VOQ_{11}$ is blocked. In this case, work conservation of input port 1 can be maintained. Both OCF and LQF achieve stability by more aggressively draining $VOQ_{11}$ than $VOQ_{12}$ in this configuration. We used this observation to propose a burst stabilization protocol that does not require comparison of state information between VOQs [28].

### 3.1 Burst stabilization protocol

A good solution to instability in VOQ switches should not require internal speed-up or the comparison of state information between VOQs. We propose that when a VOQ in an input port is selected for forwarding of a cell in the next cycle, a threshold comparison be made. As long as the current VOQ queue length exceeds a set *THRESHOLD*, then up to *BURST* cells can be transmitted from the VOQ before another VOQ from the same input port is allowed to be matched. This is similar in principle to T-RRM in [6], except in T-RRM *BURST* is effectively always 1.

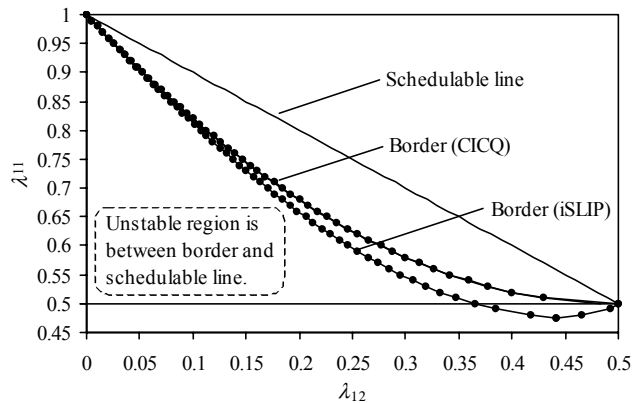

**Figure 3. Instability in RR/RR CICQ and iSLIP**

Each VOQ has a cell burst counter which decrements on consecutive cell transfers (from the VOQ). This burst counter is set to *BURST* when a VOQ drains or when the accept pointer is incremented (in iSLIP IQ) or the RR poll counter is incremented (in RR/RR CICQ). In an RR/RR CICQ switch, if a full CP buffer blocks the currently selected VOQ then the input port RR poll counter is always incremented. Specifically, for RR/RR CICQ:

- The RR poll counter in an input port is not incremented if the currently selected VOQ is above *THRESHOLD* in queue length and the cell counter is greater than zero.

The cell counter decrements on consecutive cell transfers from a VOQ. This counter is set to *BURST* when a VOQ drains or the RR poll counter is incremented (in RR/RR CICQ). If a full CP buffer blocks the currently selected VOQ, then the input port RR poll counter is always incremented. This method can also be applied to iSLIP switches [28]. For the remainder of this paper, we consider only the RR/RR CICQ switch.

### 3.2 Simulation of the burst stabilization protocol

Using our simulation model, we study the effect of *THRESHOLD* and *BURST* values on stability and delay. Fig. 4 shows mean switch delay (for $VOQ_{11}$, $VOQ_{12}$, and $VOQ_{21}$ combined) for iSLIP and RR/RR CICQ with *THRESHOLD* set to 32 and *BURST* set to 0 and 64 for $\lambda_1 = 0.99$. Also shown are results from an OCF/RR CICQ switch (the VOQs are scheduled with OCF and the CP buffers with RR). These results show that with no bursting (*BURST* = 0) instability occurs, but with bursting the switch is stable. Fig. 5 shows the mean switch delay for each VOQ for iSLIP and RR/RR CICQ switch with *THRESHOLD* and *BURST* set to 32 and 64, respectively. This shows that iSLIP and RR/RR CICQ switches with *THRESHOLD* and *BURST* have roughly similar delays for all VOQs, except $VOQ_{11}$.

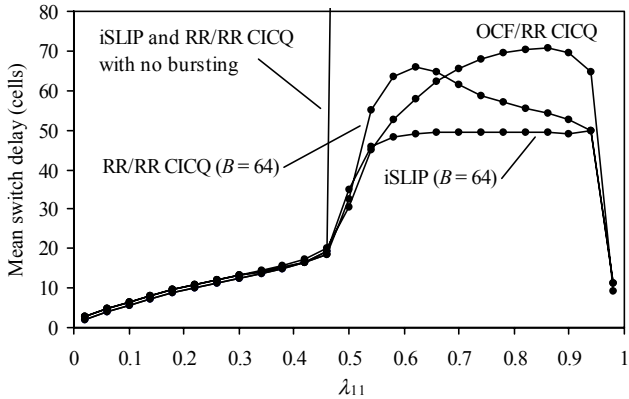

**Figure 4. Stability results for RR/RR CICQ and iSLIP**

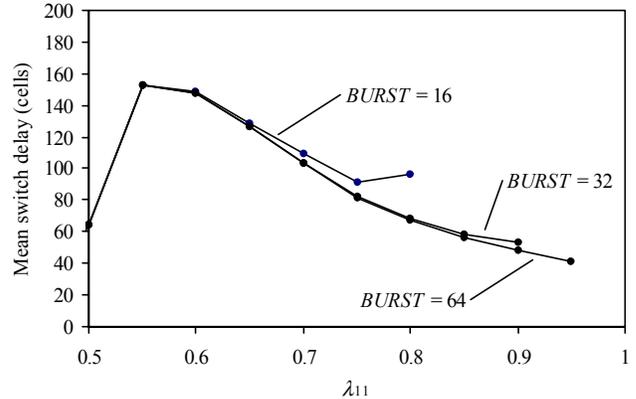

**Figure 6. Results for experiment #1**

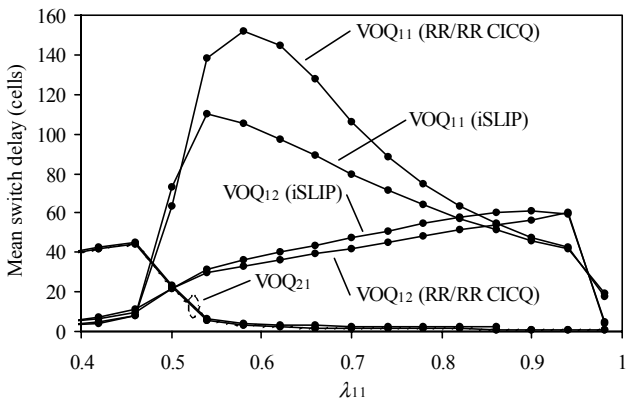

**Figure 5. Results for individual VOQs**

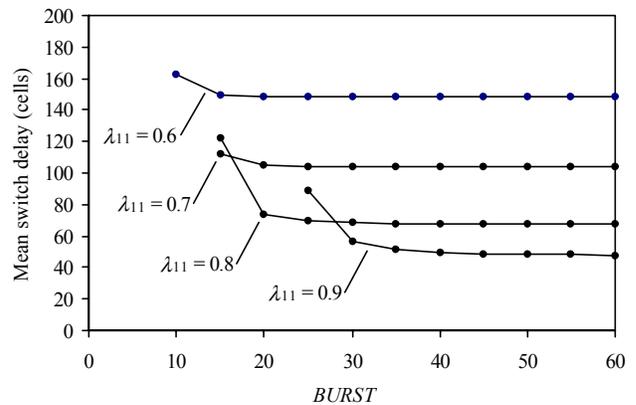

**Figure 7. Results for experiment #2**

To understand how the *THRESHOLD* and *BURST* method affects delay, three experiments varying *BURST*, *THRESHOLD*, and $\lambda_{11}$ were conducted (all with $\lambda_1 = 0.99$). The measured variable was mean switch delay. The experiments were:
- *Experiment #1:* The effect of varying $\lambda_{11}$ and *BURST* for a fixed *THRESHOLD* = 32.
- *Experiment #2:* The effect of varying *BURST* and $\lambda_{11}$ for a fixed *THRESHOLD* = 32.
- *Experiment #3:* The effect of varying *THRESHOLD* and *BURST* for a fixed $\lambda_{11} = 0.80$.

Figures 6 and 7 show the RR/RR CICQ switch mean delay for $VOQ_{11}$ for experiments #1 and #2, respectively. Figure 6 shows that mean delay for all *BURST* values are identical for $\lambda_{11} < 0.65$. Only the *BURST* value of 64 acheives stability for all $\lambda_{11}$. Figure 7 shows that larger $\lambda_{11}$ requires larger *BURST* values. These results show that a too small *BURST* value results in instability for large $\lambda_{11}$. For all cases, the mean delay for iSLIP is similar to that of RR/RR CICQ and is not shown. Figure 8 shows the mean delay for experiment #3 with *BURST* ranging from 15 to 55. *THRESHOLD* = 32 results in lower delay than *THRESHOLD* = 64, which in return has the lower delay than *THRESHOLD* = 128. A *THRESHOLD* value of 8 and 16 both result in instability. These results show that lower *THRESHOLD* values achieve lower delay, but too small of a *THRESHOLD* value results in instability. When *THRESHOLD* is too small, bursting occurs at both $VOQ_{11}$ and $VOQ_{12}$ and as a result $VOQ_{11}$ is not aggressively served (and its queue grows without bound). The mean delay for $VOQ_{12}$ decreases as the delay for VOQ$_{11}$ increases. The mean delay for VOQ$_{21}$ is not significantly affected by the values of *THRESHOLD*, thus these results are not shown in the graph.

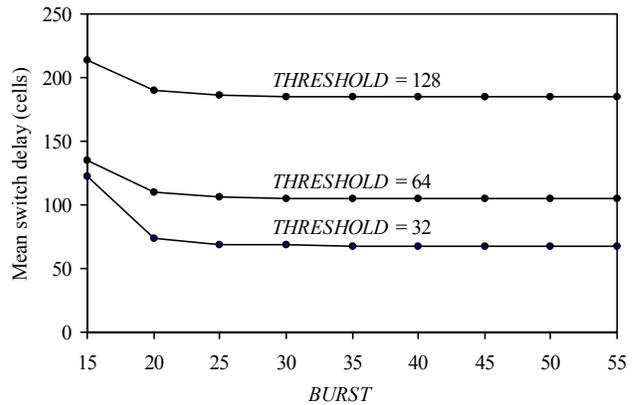

**Figure 8. Results for experiment #3**

### 3.3 An Erlang space model for unstable region

Consider each VOQ in Fig. 1 as separate queues with polling suppressed. Let $S_{ij}$ be the cell service time. The respective server utilization is then bounded by

$$\rho_{ij} = \lambda_{ij} E\{S_{ij}\} \leq 1, \quad i,j = 1,2 \qquad (1)$$

since $\lambda_{ij} \leq 1$ and $\mu^{-1} = E\{S_{ij}\} = 1$ cell time. Similarly, the total capacity $u_i$ (in Erlangs) of port $i$ is bounded $u_i \leq 1$. This capacity conservation can be used to bound the region of RR/RR CICQ stability in Erlang space ($\rho_{ij}, \rho_{ii}$) shown in Fig. 3. Although the traffic at port 1 is asymmetric $\lambda_{11} \geq \lambda_{12}$, $\rho_{11} \geq 1/2$ and $\rho_{12} \geq 1/2$ such that utilization is conserved across the two servers

$$\left(\rho_{11} - \frac{1}{2}\right) = \left(\frac{1}{2} - \rho_{12}\right) \qquad (2)$$

The trivial solution is

$$\rho_{11} = 1 - \rho_{12} \qquad (3)$$

which corresponds to the linear boundary of the unstable region in Fig. 3. Generalizing eq. (2) and noting that $\rho_{22} = 0$, capacity conservation across the servers in both ports 1 and 2 can be written as

$$\left(\rho_{11} - \frac{1}{2}\right) = \left(\frac{1}{2} - \rho_{12}\right)\left(\frac{1}{2} - \rho_{21}\right) + \left(\rho_{21} - \frac{1}{2}\right)\left(\rho_{12} - \frac{1}{2}\right) \qquad (4)$$

Since the indices can be permuted, the following simplification ensues:

$$\rho_{11} = \frac{1}{2} + 2\left(\frac{1}{2} - \rho_{12}\right)\left(\frac{1}{2} - \rho_{21}\right)$$

$$= \frac{1}{2} + 2\left(\frac{1}{2} - \rho_{12}\right)^2 \qquad (5)$$

$$= 1 - 2\rho_{12} + 2\rho_{12}^2 \qquad (6)$$

Eq. (5) is recognizable as a conic section with eccentricity $e = 1$ and vertex $\{h,k\} = (1/2, 1/2)$ and corresponds to the parabolic locus in Fig. 3. Both eq. (3) and eq. (6) confirm the alternative derivation in [28].

## 4. An Analytical Model of Burst Stabilization

In this section we develop an analytical model to predict the minimum *BURST* value needed to stabilize the system. We want to establish two things concerning the burst stabilization protocol:
1. That it is sufficient to inhibit such instabilities. This is discussed in Section 4.1.
2. That the magnitude of the *BURST* parameter can be predicted with sufficient accuracy for the important range of loads expected in a real switch. Section 4.5 contains those details

The polled queues at each port are intrinsically stable if inputs and outputs are not oversubscribed. That is,

$$\sum_{i=1}^{N} \lambda_{ij} \leq 1, \quad \forall i < j$$

The instability (described in Section 3) that we are concerned with arises primarily from the blocking of a VOQ on one port by the transmission from a VOQ on another port via the corresponding output CP buffer on the crossbar. For example, buffer CP$_{ii}$ blocks transmissions from VOQ$_{ii}$ because of the presence of a cell in the downstream buffer CP$_{ji}$ due to transmissions from VOQ$_{ji}$ where $j > i$.

We show that $\hat{B}_{ij}$, the estimate for the minimum *BURST* parameter, comprises two terms:

$$\hat{B}_{ij} = B_i + B_j \qquad (7)$$

where $B_i$ is due to traffic arriving at port $i$ and $B_j$ due to traffic arriving at port $j > i$. The interaction between these two traffic sources is such that their contribution to the minimum *BURST* size is both additive and load-dependent. In the subsequent discussion, $\hat{B}_{ij}$ signifies the minimum value of the *BURST* parameter required to stabilize the VOQs. All queue lengths are defined with respect to the *THRESHOLD* value which (as mentioned in Section 3.3) acts as an arbitrary reference level.

Stability analysis is intrinsically difficult because transient effects [7, 12, 13, 17] may not possess a closed analytic form and only asymptotic bounds may be represented [10, 25]. Moreover, the VOQs at each input port in Fig. 1 are subject to asymmetric traffic and even the steady-state behavior of such asymmetric polling systems can be difficult to express analytically [22].

Surprisingly, however, we present an accurate steady-state bound for RR/RR CICQ switch stabilization based on the fact that certain aspects of our problem resemble the equilibrium queue length

$$E\{Q\} = \frac{\lambda^2}{\mu-\lambda}\frac{(C_s^2-1)}{2\mu} + \frac{\lambda}{\mu-\lambda} \quad (8)$$

for an M/G/1 model of an exhaustive polling system [2, 3, 14] with $C_s^2$ the squared coefficient of variation of the service time $S$.

### 4.1 Vacating server approximation

The principles of operation of the burst stabilization protocol are best understood in the context of a simplified model having a single burst-stabilized queue. The generalization from this simple burst model to the multi-queue configuration in the real RR/RR CICQ switch proceeds in a straightforward way.

Consider a single queue with arrivals that are Poisson distributed with rate $\lambda \geq 1/2$ and serviced in FCFS order. Poisson-distributed events have interarrival periods that are exponentially distributed and the latter distribution is the continuous analog of the discrete Bernoulli distribution used in the simulations of Section 3. Upon servicing a single request, the server vacates the queueing center for one service period $E\{S\} = \mu^{-1}$. During that vacation period, other requests may arrive into the queue. From the standpoint of an arriving request, the expected service time appears to be $E\{S\} = 2$ because processing time is split equally between servicing the next request at the head of the queue and the next vacation period i.e., an effective rate $\mu = 1/2$. Since $\lambda \geq \mu$, such a queue is non-ergodic and therefore subject to unstable queue growth.

Now, suppose that when the queue size exceeds *THRESHOLD* the server ceases vacating the queue and proceeds to service at most *BURST* requests at a rate $\mu = 1$ before taking the next vacation period. Is it possible to find a *BURST* value large enough to bound the queue size over a sufficiently long time period? We could attempt to address this question using an M/G/1 generalized service time distribution model [2], however, it will prove more instructive for our later discussion to present a simpler rate matching argument.

In a fluid approximation [7, 12] the necessary condition (see Proof 1 in the appendix) for stability is that *BURST* (*B*) requests be serviced in a time $\tau$ such that:

$$\frac{B}{\tau} = \lambda \quad (9)$$

The time-averaged rate of service must match the mean arrival rate in the long run. Applying this condition to our vacating server model, at most *B* requests must be serviced in a period $\tau = B+1$; where the "1" refers to the mean vacation period. The rate equation,

$$\frac{B\mu}{B+1} = \lambda \quad (10)$$

corresponds to the fraction of time for which the effective service rate reaches $\mu = 1$ in this model. It follows that the *BURST* size must be

$$B \geq \frac{\lambda}{\mu-\lambda} \quad (11)$$

which is finite and bounded provided $\lambda < \mu$.

It is noteworthy that a term similar to eq. (11) arises in the steady-state limit of the rate processing function for leaky bucket queue management [9]. More importantly for our discussion, it corresponds to the second term in eq. (8). We capitalize on this observation in Section 4.3.

### 4.2 Port 2 analysis

Continuing this line of thought, it turns out to be easier to understand the contribution to *BURST* from port $j=2$ interactions before turning to those due to port $i=1$ traffic.

**Definition 2** *Let $\lambda_2$ be the offered traffic at port 2. The fraction of offered traffic going to VOQ$_{21}$ is $\lambda_{21} = \lambda_{12} + (1-\lambda_1)$ at instability and $\lambda_{22} = 0$.*

**Corollary 2** $\lambda_{12} = \lambda_{21}$ iff $\lambda_1 = 1$.

The mean interarrival time of cells at port 2 is $\tau_2 = \lambda_2^{-1}$. But $\lambda_{21} \equiv \lambda_2$ by virtue of $\lambda_{22} = 0$. Then $\tau_2 = \tau_{21} = \lambda_{21}^{-1}$, and $\tau_{21} \equiv \lambda_{21}^{-1} = 1 + B_2$ by analogy with the discussion in Section 4.1. Applying Proof 2 (see the appendix), the contribution to *BURST* from port 2 is:

$$B_2 = \frac{1-\lambda_{21}}{\lambda_{21}} \equiv \frac{\lambda_{11}}{1-\lambda_{11}} \quad (12)$$

Finally, eq. (12) can be rewritten in terms of the fraction of traffic going to VOQ$_{11}$ as

$$B_2(f) = \frac{f\lambda_1}{1-f\lambda_1} \quad (13)$$

by application of Definition 1.

This result states that the mean number of cells burst from VOQ$_{11}$ prior to processing being blocked by transmissions from VOQ$_{12}$ is the same as the equilibrium queue size for the vacating server model in Section 4.1. The reason is that a cell will be present at VOQ$_{21}$ every $\tau_{21}$ cell times (on average) causing the scheduler to vacate VOQ$_{11}$ and service VOQ$_{12}$.

### 4.3 Port 1 analysis

We now turn to the analysis of port $i = 1$. Because bursts can be interrupted with a mean time $\tau_{21}$, there is also the possibility that VOQ$_{12}$ can burst. As seen from the standpoint of VOQ$_{11}$ the vacation period is extended beyond that accounted for by eq. (13). Since cells will continue to arrive into VOQ$_{11}$ the burst size will need to be larger than $B_2$.

The average number of arriving cells $L_1$ that accumulate during this extended vacation period is directly proportional to the arrival rate at port 1 according to Little's Law,

$$L_1 = \lambda_1 W_1 \qquad (14)$$

where $W_1$ is the expected waiting-time. Eq. (14) corresponds to the first term in eq. (8) viz,

$$L_1 = \frac{\lambda_1^2}{1-\lambda_1} \frac{(C_s^2 - 1)}{2} \qquad (15)$$

with $\mu = 1$. In polling systems near saturation, waiting-times approach a gamma distribution and higher moment effects vanish under heavy traffic [24].

The instability of interest to us, however, arises from heavy asymmetric traffic into VOQ$_{11}$. This has the effect of making the residual service time in eq. (15) a load-dependent function via the squared coefficient of variation

$$C_s^2(f) = 1 + \frac{4}{5}\left(f - \frac{1}{2}\right) \qquad (16)$$

such that $B_1$ in eq. (7) becomes:

$$B_1(f) = \frac{2}{5}\left(\frac{\lambda_1^2}{1-\lambda_1}\right)\left(f - \frac{1}{2}\right) \qquad (17)$$

with $f \geq 1/2$ the asymmetric fraction of the offered load arriving at VOQ$_{11}$. Note that $f$ is the only variable in eq. (17).

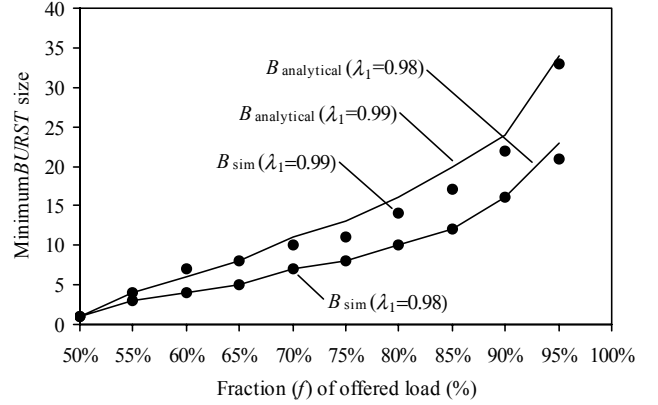

**Figure 9. Prediction of minimum *BURST* value**

As expected the length of the waiting line, and hence its contribution to the minimum *BURST* size, is directly proportional to the imbalance in VOQ$_{11}$ traffic. When $f = 1/2$ (symmetric case) this term vanishes and does not contribute $\hat{B}_{ij}$.

### 4.4 Bounds on *BURST*

The complete expression for estimating the minimum *BURST* as defined in eq. (7) is:

$$\hat{B}_{ij}(f) = \frac{2}{5}\left(\frac{\lambda_i^2}{1-\lambda_i}\right)\left(f - \frac{1}{2}\right) + \frac{f\lambda_i}{1-f\lambda_i}, \quad \forall i < j \qquad (18)$$

More formally *BURST* corresponds to the infimum (greatest lower bound) of $\hat{B}_{ij}(f)$. As a practical matter, this can be evaluated most simply as the ceiling function $\lceil \cdot \rceil$ applied to eq. (18). For $\lambda_i \leq 1$ and $f = 1/2$ we have

- $\left\lceil B_i\left(\frac{1}{2}\right) \right\rceil = 1$
- $\left\lceil B_j\left(\frac{1}{2}\right) \right\rceil = 0$

As expected from the simulation experiments (Fig. 9), eq. (18) is an increasing function of $f$ but not necessarily monotonic. It does not scale with the number of ports ($N$) since blocking between any pair of ports (as remarked in Section 3) is sufficient to cause unstable queue growth. Eq. (18) is rather remarkable in that it is based on a steady-state representation yet the fluctuations about these mean values can be very significant. It is also surprising that it can be expressed entirely in terms of the fractional load at VOQ$_{ii}$. It should be noted that eq. (18) is not valid for $f = 1$ since there is a cell in every slot and the behavior becomes D/D/1 (i.e., non-Poisson arrivals).

**Table 1. Calculated and simulated minimum BURST values for $\lambda_1 = 0.98$**

| Loads | | | Model | | | Comparison | | |
|---|---|---|---|---|---|---|---|---|
| $f$ | $\lambda_{11}$ | $\lambda_{12}$ | $B_2$ | $B_1$ | $B_{12}$ | $\lceil \hat{B}_{12} \rceil$ | $B_{sim}$ | Error |
| 0.50 | 0.49 | 0.49 | 0.96 | 0.00 | 0.96 | 1 | 1 | 0.00 |
| 0.55 | 0.54 | 0.44 | 1.17 | 0.96 | 2.13 | 3 | 3 | 0.00 |
| 0.60 | 0.59 | 0.39 | 1.43 | 1.92 | 3.35 | 4 | 4 | 0.00 |
| 0.65 | 0.64 | 0.34 | 1.75 | 2.88 | 4.64 | 5 | 5 | 0.00 |
| 0.70 | 0.69 | 0.29 | 2.18 | 3.84 | 6.03 | 7 | 7 | 0.00 |
| 0.75 | 0.74 | 0.25 | 2.77 | 4.80 | 7.58 | 8 | 8 | 0.00 |
| 0.80 | 0.78 | 0.20 | 3.63 | 5.76 | 9.39 | 10 | 10 | 0.00 |
| 0.85 | 0.83 | 0.15 | 4.99 | 6.72 | 11.71 | 12 | 12 | 0.00 |
| 0.90 | 0.88 | 0.10 | 7.47 | 7.68 | 15.16 | 16 | 16 | 0.00 |
| 0.95 | 0.93 | 0.05 | 13.49 | 8.64 | 22.14 | 23 | 21 | + 0.10 |

**Table 2. Calculated and simulated minimum BURST values for $\lambda_1 = 0.99$**

| Loads | | | Model | | | Comparison | | |
|---|---|---|---|---|---|---|---|---|
| $f$ | $\lambda_{11}$ | $\lambda_{12}$ | $B_2$ | $B_1$ | $B_{12}$ | $\lceil \hat{B}_{12} \rceil$ | $B_{sim}$ | Error |
| 0.50 | 0.54 | 0.45 | 0.98 | 0.00 | 0.98 | 1 | 1 | 0.00 |
| 0.55 | 0.54 | 0.45 | 1.20 | 1.96 | 3.16 | 4 | 4 | 0.00 |
| 0.60 | 0.59 | 0.40 | 1.46 | 3.92 | 5.38 | 6 | 7 | − 0.14 |
| 0.65 | 0.64 | 0.35 | 1.81 | 5.88 | 7.69 | 8 | 8 | 0.00 |
| 0.70 | 0.69 | 0.30 | 2.26 | 7.84 | 10.10 | 11 | 10 | + 0.10 |
| 0.75 | 0.74 | 0.25 | 2.88 | 9.80 | 12.68 | 13 | 11 | + 0.18 |
| 0.80 | 0.79 | 0.20 | 3.81 | 11.76 | 15.57 | 16 | 14 | + 0.14 |
| 0.85 | 0.84 | 0.15 | 5.31 | 13.72 | 19.03 | 20 | 17 | + 0.18 |
| 0.90 | 0.89 | 0.10 | 8.17 | 15.68 | 23.86 | 24 | 22 | + 0.09 |
| 0.95 | 0.94 | 0.05 | 15.81 | 17.64 | 33.45 | 34 | 33 | + 0.03 |

### 4.5 Numerical results

For the simulation results, the experiment of Section 3.2 was repeated (again using *THRESHOLD* = 32) with a *BURST* size that was incrementally increased across each run until the sampled drift in VOQ$_{11}$ queue growth was observed to be adiabatically zero. Tables 1 and 2 present the results for a range of $\lambda_1$ values and a comparison with the corresponding predictions of eq. (18). Under- and over-estimations are indicated respectively by (–) and (+) signs. Fig. 9 also shows the measured ($B_{sim}$) and the predicted ($B_{analytical}$) minimum *BURST* values. $B_{analytical}$ is defined to be equivalent to $\lceil \hat{B}_{12} \rceil$. These results demonstrate that increasing the value of *BURST* achieves stability by increasing the bandwidth for VOQ$_{11}$ at the expense of VOQ$_{12}$. For $\lambda_1 = 0.98$ and $\lambda_1 = 0.99$ we determine the value of *BURST* for which stability is achieved. In this way we establish that eq. (18) predicts the exact *BURST* size with less than or equal to 10% relative error for $\lambda_1 = 0.98$ and less than 20% relative error for $\lambda_1 = 0.99$. In all cases, except one, the relative errors are conservative since they are overestimates.

## 5. Summary and Future Work

We have proposed and investigated a burst stabilization protocol to make iSLIP IQ and RR/RR CICQ switches stable for the unstable region identified in [14]. This unstable region occurs when asymmetric arrivals occur at any two input ports in a switch. The new protocol uses a queue length threshold in the switch VOQ buffers. When a *THRESHOLD* value is exceeded, the VOQ is allowed to send up to a *BURST* number of cells sequentially. The *THRESHOLD* value is configured into the switch and, as shown in our simulation results, a *THRESHOLD* of greater than about 32 cells is sufficient for stability.

We have shown how the burst stabilization protocol stabilizes the VOQs under heavy traffic. The dynamics of this protocol is complex but is most easily understood as a subtle extension of an M/G/1 polling system with vacations. The subtleties arise from the superposition of two dominant effects: i) the vacation periods are caused by pairwise interference between VOQs on different ports and ii) the waiting time at the busiest of the two ports is the result of a residual service time that is dependent on the magnitude of the asymmetric load. This extended M/G/1 model is sufficient to predict the minimum *BURST* size required for VOQ stability in the presence of unbalanced traffic.

Given the continuing increase in VLSI density, we believe that RR/RR CICQ switches hold great promise for the future [28] and that the results in this paper are thus very significant in showing the feasibility of achieving stability in this architecture without requiring internal speed-up. As part of our future work we would like to investigate further if it is possible to derive this equilibrium model from the transient dynamics seen in the simulations. We would also like to extend the analytical model for iSLIP.

## Appendix – Proofs for Stabilization Model

**Proof 1** *In the notation of [7] and [12], the discrete time representation of the sample path representing the queue length is*

$$Z(n) = Z(0) + A(n) - D(n) \tag{19}$$

*where $Z(n)$ corresponds to the number of cells in the queue at the commencement of discrete time-slot n. $A(n)$ corresponds to the cell count that have arrived into the queue by time n, $D(n)$ corresponds to the cell count that have been serviced by time n, and port subscripts (i, j) are redundant for the single queue vacating server model of Section 4.1.*

*Let l be the arrival rate in the queue and assume that*

$$\lim_{n \to \infty} \frac{A(n)}{n} = \lambda \tag{20}$$

*Similarly, the queue will be rate stable with probability one if,*

$$\lim_{n \to \infty} \frac{D(n)}{n} = \lambda \tag{21}$$

*See [7] for the complete proof.*

*The equivalent fluid approximation is a continuous-time t representation where eq. (19) is replaced by*

$$E\{Z(t)\} = E\{Z(0)\} + \lambda t - E\{D(t)\} \geq 0 \tag{22}$$

*with $E\{\cdot\}$ the time-averaged mean value. In the notation of Section 4.1, $E\{Z(t)\} \equiv E\{Q(t)\}$.*

*Assuming the time derivative exists, the condition for stability is that the mean drift function*

$$\dot{Q}(t) = \lambda - \dot{D}(t) \tag{23}$$

*vanish in the long run i.e.,*

$$\lim_{t \to \infty} \frac{D(t)}{t} = \lambda \tag{24}$$

*which corresponds to eq. (9).*

**Proof 2** *The mean time ($\lambda_{21}$) between $CP_{12}$ blocking $CP_{11}$ via arrivals at port 2 is related to the mean interarrival rate ($\lambda_{21}$) at $VOQ_{21}$ by:*

$$\lambda_{21}^{-1} \equiv \tau_{21} = B_2 + 1$$

*Applying Definition 2 and rearranging terms as follows:*

$$B_2 = \frac{1}{\lambda_{12} + (1 - \lambda_1)} - 1$$

$$= \frac{\lambda_1 - \lambda_{12}}{1 - \lambda_1 + \lambda_{12}}$$

*Corollary 1 affords the simplification:*

$$B_2 = \frac{\lambda_{11}}{1 - \lambda_{11}}$$

*and eq. (13) follows by virtue of Definition 1.*

## Acknowledgment

The authors thank the anonymous referees for their helpful suggestions.